\documentclass[useAMS, usenatbib]{mn2e}

\usepackage[dvips]{graphicx,color}
\usepackage[]{txfonts}

\def\simless{\mathbin{\lower 3pt\hbox
{$\rlap{\raise 5pt\hbox{$\char'074$}}\mathchar"7218$}}}   
\def\simmore{\mathbin{\lower 3pt\hbox
{$\rlap{\raise 5pt\hbox{$\char'076$}}\mathchar"7218$}}}   
\newcommand{\dsfrac}[2]{\displaystyle{\frac{#1}{#2}}}

\newcommand{\secref}[1]{Section~\ref{#1}}
\newcommand{\figref}[1]{Fig.~\ref{#1}}

\title[Afterglows from magnetized GRB flows]{Multiwavelength
afterglow light curves from magnetized GRB flows}
\author[P. Mimica, D. Giannios and
M. A. Aloy]{P. Mimica$^{1}$\thanks{E-mail: Petar.Mimica@uv.es}, D. Giannios$^{2}$ and M. A. Aloy$^{1}$\\
$^{1}$Departamento de Astronom\'ia y Astrof\'isica, Universidad de
Valencia, 46100, Burjassot, Spain\\
$^{2}$Department of Astrophysical Sciences, Peyton Hall, Princeton University, Princeton, NJ 08544, USA}

\begin{document}

\maketitle

\label{firstpage}

\begin{abstract}
  We use high-resolution relativistic MHD simulations coupled with a
  radiative transfer code to compute multiwavelength afterglow light
  curves of magnetized ejecta of gamma-ray bursts interacting with a
  uniform circumburst medium. The aim of our study is to determine how
  the magnetization of the ejecta at large distance from the central
  engine influences the afterglow emission, and to assess whether
  observations can be reliably used to infer the strength of the
  magnetic field. We find that, for typical parameters of the ejecta,
  the emission from the reverse shock peaks for magnetization
  $\sigma_0\sim 0.01-0.1$ of the flow, and that it is greatly suppressed
  for higher $\sigma_0$.  The emission from the forward shock shows an
  achromatic break shortly after the end of the burst marking the
  onset of the self-similar evolution of the blast wave. Fitting the
  early afterglow of GRB~990123 and 090102 with our numerical models
  we infer respective magnetizations of $\sigma_0\sim 0.01$ and
  $\sigma_0\sim 0.1$ for these bursts. We argue that the lack of
  observed reverse shock emission from the majority of the bursts can
  be understood if $\sigma_0\simmore 0.1$, since we obtain that the
  luminosity of the reverse shock decreases significantly for
  $\sigma_0\sim 1$. For ejecta with $\sigma_0\simmore 0.1$ our models
  predict that there is sufficient energy left in the magnetic field,
  at least during an interval of $\sim 10$ times the burst duration,
  to produce a substantial emission if the magnetic energy can be
  dissipated (for instance, due to resistive effects) and radiated
  away.
\end{abstract}


\section{Introduction}

Gamma ray bursts (GRBs) are thought to be caused by the energy
dissipated by an ultra-relativistic outflow. Two alternative scenarios
for powering the flow have been extensively considered, resulting
either in the thermal-energy dominated fireball
\citep{Goodman:1986eq,Paczynski:1986qd} or in the Poynting-flux
dominated flow
\citep[PDF:][]{Usov:1992hp,Thompson:1994ft,Meszaros:1997bx}. In the
fireball model, magnetic fields are not dynamically important, and at
large distances the flow is weakly magnetized $\sigma_0 \ll
1$. Radiation pressure drives the acceleration of the flow until the
bulk Lorentz factor becomes $\Gamma_0 \simeq h_{\rm in}$ \citep[where
$h_{\rm in}$ is the initial specific enthalphy of the
flow; see e.g.,][]{Aloy:2005zp,Aloy:2007p7001}, and internal collisions power the
prompt emission \citep{Rees:1994ca}. MHD models for GRB jets assume
that the flow is launched Poynting-flux dominated ($\sigma_{\rm in}\gg
1$; just as in the widely accepted theoretical model for relativistic
jets from active galactic nuclei). The acceleration of the flow is
model dependent and therefore uncertain. PDF flows have in common the
fact that the acceleration to ultra-relativistic speed happens by
converting part of the magnetic energy into kinetic energy
\citep[e.g.,][]{Li:1992ez}. However, there is no full consensus on
which is the magnetization of the flow ($\sigma_0$) when the
acceleration is completed. \citet{Lyutikov:2003lk} consider a sub-fast magnetosonic flow
that maintains $\sigma_0\gg 1$ up to distances of $r> 10^{16}\,$cm,
where it interacts with the surrounding medium. In
contrast, a large number of ideal and non-ideal MHD models
for jet acceleration are super-fast magnetosonic. Still, they support a picture where the
acceleration/dissipation processes are not 100\% efficient in
consuming the magnetic energy and, therefore, the flow magnetization
$\sigma_0\sim 1$ and the bulk Lorentz factor is $\Gamma_0\simeq
\sigma_{\rm in}/(1+\sigma_0)\sim \sigma_{\rm in}$ at large distances
\citep{Drenkhahn:2002zm, Komissarov:2009p2165, Tchekhovskoy:2009p7068,
  Lyubarsky:2010p3273}.  During or after the acceleration process
magnetic dissipation
\citep{Thompson:1994ft,Spruit:2001hi,Giannios:2008p7077} or internal
shocks \citep{Fan:2004p1007,Mimica:2010p6852} may power the GRB.  Here
we focus on the dynamics of the deceleration of $\sigma_0 \simless 1$
flows with focus on the role of magnetic field strength to the
resulting emission.

If the magnetic fields of the ejecta are dynamically unimportant
(e.g., $\sigma_0\ll 1$) at large distance where the interaction with the
circumburst medium becomes substantial, a forward shock (FS) and a
reverse shock (RS) are expected to form at the interface between the
GRB ejecta and circumburst (external) medium \citep{Sari:1995oq}. If
the ejecta is strongly magnetized, we expect a weak or absent RS
\citep{Giannios:2008zl}. Observationally, there is a wealth of recent
evidence that magnetic fields shall be dynamically important. Most
bursts show no trace of the RS emission, in contrast to predictions of
the fireball model \citep[see, however,
][]{Nakar:2005p4038,McMahon:2006p4021}.  When the bump in the early
optical light curve is observed (and resembles that expected by the
RS), modeling of the light curves indicates that the magnetic field is
larger in the shocked ejecta than in the shocked external medium
\citep{Fan:2002p7207,Zhang:2003wc,Panaitescu:2004p7344}.  More
conclusive is the detection of $\sim 10\%$ polarization in the early
optical afterglow emission of GRB 090102, consistent with the ejecta
containing quasi-coherent, large-scale magnetic fields
\citep{Steele:2009p5925}. Since dynamically dominant magnetic fields
would suppress the RS emission altogether, these authors give a rough
estimate of $\sigma_0 \sim 1$ for GRB 090102, supporting super-fast
magnetosonic MHD models for GRBs.

To make quantified inferences for the degree of the magnetization of
the ejecta, a detailed understanding of the dynamical interaction
between magnetized ejecta and external medium is needed. This has been
done at an increasing level of sophistication. As a first approach to
the problem, the dynamical effects of magnetic fields have been
ignored, assuming a purely hydrodynamic behavior of the ejecta
\citep{Rees:1992kx,Sari:1995oq,Kobayashi:1999ly,Granot:2001fq,Meliani:2007qm}.
Such an approximation has lead to infer that the magnetic field
strength at the RS and at the FS is different (i.e., the parameter
$\epsilon_{B}$ is larger at the RS than at the FS
\citep{Fan:2002p7207,Zhang:2003wc}).  A more
accurate approach takes into account the modification in the shock
jump conditions because of the presence of magnetic fields
\citep{Fan:2004if,Zhang:2005ts}.  \citet{Lyutikov:2006wj,Genet:2006fe}
studied the afterglow from sub-fast ($\sigma_0\gg 1$) MHD models.
\citet{Giannios:2008zl} found that rarefaction waves, instead of
shocks, can also form in magnetized ejecta \citep[see
also][]{Mizuno:2009p210,Aloy:2008kx}. These studies have left several issues open
(e.g., that of the timescale of transfer of the magnetic energy into
the shocked external medium) or debated
\citep[see][]{Lyutikov:2005tx}.  The complete dynamical evolution of
homogeneous, magnetized ejecta colliding with the ISM has been
simulated by \citep[][MGA09 in the following]{Mimica:2009qa} for a few
characteristic cases, showing how the magnetic energy is transferred
in the shocked ISM until a self-similar blast wave forms at later
time.

Although bolometric light curves of the expected emission of
magnetized ejecta were computed by MGA09, a detailed calculation of
the synthetic light curves at different wavelengths is missing. In this
work we have coupled the SPEV (\emph{SP}ectral \emph{EV}olution) code
\citep{Mimica:2009p1445}, used to compute the non-thermal emission
from relativistic outflows, to the MRGENESIS
\citep{Mimica:2005sp,Mimica:2007db} RMHD code in order to study in
accurately the signatures of magnetic fields and the (non)existence of
shocks in afterglow ejecta.

The paper is organized as follows: in \secref{sec:numaftsim} we give a
brief overview of the results in MGA09 and present our extended set of
models. \secref{sec:non-thermal} gives details of the model we use to
compute the non-thermal synchrotron emission from relativistic
shocks. In \secref{sec:dynamics} we discuss the energy content of GRB
ejecta and of the external medium as a function of the distance and of
the initial flow magnetization. Light curves and spectra are presented
in \secref{sec:lightcurves}. We apply our models to the GRBs 990123
and 090102 in \secref{sec:application}. Discussion and conclusions are
given in \secref{sec:discussion}.

\section{Numerical modeling of ejecta-medium interaction}
\label{sec:numaftsim}

We are interested in following the evolution of the flow well after
the collimation, acceleration, and GRB emission phases. At this stage,
the ejecta are moving radially, cold (because of adiabatic expansion)
and expected to contain a dominant toroidal component of the magnetic
field.  We parametrize the magnetization of the ejecta through
$\sigma_0 \equiv B'^2_{\rm mac}/4\pi \rho c^2$, where $B'_{\rm mac}$ and
$\rho$ are the rest-frame macroscopic (ordered) magnetic field
strength and density respectively.

We focus on the initial stages of the interactions during which only
moderate deceleration of the ejecta takes place. The ejecta are
expected to have bulk Lorenz factor $\Gamma_0\gg 1/\theta$, where
$\theta$ is opening angle of the jet. This allows us to ignore 2D
effects (such as lateral jet spreading) justifying our assumed
spherical symmetry.  The ejecta is modeled as a homogeneous shell of
initial thickness $\Delta_0\ll r_0$, where $r_0$ is the initial
distance determined by the following considerations. Because of its
magnetic pressure, the shell will tend to expand developing
rarefactions on both radial edges. These rarefactions propagate at
fast magnetosonic speed in the ejecta. Comparing the time it takes for
a rarefaction to cross the ejecta with the expansion timescale,
\citet{Giannios:2008zl} estimate the distance at which the
rarefactions broaden the shell. Furthermore, a reverse shock may form
in the ejecta as result of the interaction with the ISM, or as the
late steepening of the conditions in the tail of the primordial
rarefaction sweeping backwards the ejecta from its leading radial edge
(MGA09). We set the initial distance $r_0$ of our study such that it
is large enough to make the study amenable to numerical RMHD
simulations, but small enough so as to begin our simulations in the
coasting phase of the ejecta, before the deceleration of the
ultra-relativistic flow. Hence, we can track self-consistently, both
the evolution of the complete Riemann fan emerging from each radial
edge of the shell and the initial deceleration of the ejecta.

The single quantity that determines the strength of the reverse shock
in unmagnetized ejecta is $\xi\equiv \sqrt{l/\Delta_0}\Gamma_0^{-4/3}$
where $l=(3E/4\pi n_{\rm ext} m_{\rm p}c^2)^{1/3}$ is the Sedov
length, $n_{\rm ext}$ is the number density of the external medium,
$c$ is the speed of light, $m_p$ is the proton mass, and $E=E_k=4\pi
r^2_0\Delta_0\Gamma^2_0 \rho_{\rm shell} c^2$ is the total (kinetic,
$E_k$) energy of the ejecta (the internal energy contribution is
neglected, since the shell is initially assumed to be cold). The $\xi$
quantity can be related to the more familiar ``deceleration radius''
$r_{\rm dec}$ defined as the radius where the ejecta accumulate a mass
$\Gamma_0^{-1}$ times their own mass from the external medium,
resulting in $r_{\rm dec}=l/\Gamma_0^{2/3}$. For parameters relevant
for GRB flows, the parameter $\xi$ is expected to be of order of unity
(within one order of magnitude).  The second parameter that naturally
appears when studying the deceleration of magnetized ejecta is the
magnetization $\sigma_0$ in the coasting phase ($\sigma_0$ is much
less constrained from theory or observations). We note that in
magnetized ejecta, the total (kinetic plus magnetic) energy is $E=E_k
(1 + \sigma_0)$. Realistic GRB models involve ejecta initially moving
with Lorenz factor $\Gamma_0 \sim 100-1000$ and with thickness
$\Delta_0 \sim c T_{\rm GRB}\sim 10^{12} {\rm cm}\ll r_{\rm dec}$
($T_{\rm GRB}$ is the duration of the GRB phase). Such simulations
pose large numerical challenges. In MGA09 we overcome these challenges
by showing that one can simulate a relativistically moving shell of,
say $\Gamma_0\sim$ tens and {\it predict} the evolution of a shell
with larger $\Gamma_0$ but with the same $\sigma_0$ and $\xi$.  We
refer to this procedure as rescaling.

In this work, we extend the models of MGA09 to explore a wider
parameter space of $\sigma_0$ for two values of $\xi=0.5$ and
$\xi=1.1$. We refer to the $\xi=0.5$ runs as ``thick shell'' and
$\xi=1.1$ as ``thin shell'' models respectively. In Tab.~\ref{table:1}
we summarize parameters of the numerical models. $E$, $\sigma_0$, $r_0$,
$\Delta_0$ and $\Gamma_0$ are the shell total energy, magnetization,
initial radius, width and initial Lorenz factor, respectively. We
assume that both the external medium and the ejecta are ideal gases
which obey an equation of state with a fixed adiabatic index
$\gamma_{\rm ad}=4/3$, that $\rho = n m_p$ and give in
Tab.~\ref{table:1} the comoving number density of the external medium
$n_{\rm ext}$, and the pressure to rest-mass density ratios $P_{\rm
  ext}/\rho_{\rm ext}c^2$ and $P_{\rm shell}/\rho_{\rm shell}c^2$ in
the external medium and in the shell, respectively. Finally, we
indicate the Lorenz factor $\Gamma_{0, {\rm rescaled}}$ and shell
thickness $\Delta_{0, \rm rescaled}$ to which results of our
simulations are rescaled (see \secref{sec:lightcurves}).

\begin{table}
{\large
  \caption{Properties of the simulated ejecta-external medium interactions. See text of
    Sect.~\ref{sec:numaftsim} for the definitions
    of parameters.}
\label{table:1}
\centering 
\begin{tabular}{lcc}
  \hline
  & thin shell ($\xi = 1.1$) & thick shell ($\xi = 0.5$) \\
  \hline
  $E$ [erg] & $3.33\times 10^{53}$ & $3.33\times 10^{54}$\\
  $\sigma_0$ & \multicolumn{2}{|c|}{$0$, $\:0.0001$, $\:0.001$, $\:0.01$,
    $\:0.05$, $\:0.1$, $\:0.5$, $\:1$}\\
  $n_{\rm ext}$ [cm$^{-3}$] & \multicolumn{2}{|c|}{$10$}\\
  $P_{\rm ext}/\rho_{\rm ext}c^2$ & \multicolumn{2}{|c|}{$10^{-3}$}\\ 
  $P_{\rm shell}/\rho_{\rm shell}c^2$ &  \multicolumn{2}{|c|}{$2.5\times 10^{-3}$}\\ 
  $r_0$ [cm] & \multicolumn{2}{|c|}{$5\times 10^{16}$}\\
  $r_{0,{\rm rescaled}}$ [cm] & {$6.8\times 10^{15}$} & {$3.7\times 10^{15}$}\\
  $\Delta_0$ [cm] & $10^{15}$ & $10^{16}$\\
  $\Delta_{0,{\rm rescaled}}$ [cm] & $3.4\times 10^{11}$ & $3\times 10^{11}$\\
  $\Gamma_0$ & \multicolumn{2}{|c|}{$15$}\\
  $\Gamma_{0, {\rm rescaled}}$ & $300$ & $750$\\
\hline
\end{tabular}
}
\end{table}

For the numerical simulations we use the relativistic
magnetohydrodynamic code \emph{MRGENESIS}
\citep{Mimica:2005sp,Mimica:2007db}, a high-resolution shock capturing
scheme based on \emph{GENESIS} \citep{Aloy:1999rm,Leismann:2005rz}. In
the problem at hand, we assume spherical symmetry and the fluid is
discretized in spherical shells (zones).  As described in detail in
MGA09, we use the grid re-mapping technique, which enables us to
decrease the computational cost of our simulations by concentrating
the computation on a domain, which is several initial shell widths
wide, and which follows the shell as it moves radially outwards. Since
the disadvantage of this is that at later times we lose a part of the
tail of the ejecta, we have recomputed some of the models listed in
Tab.~\ref{table:1} with a five times wider radial grid. Thus, we can
correctly follow the evolution of the total magnetic energy content to
later times (see \secref{sec:longterm}).

\section{Non-thermal emission}
\label{sec:non-thermal}

The bulk of the observed radiation that follows the prompt GRB phase
is believed to be synchrotron emission from electrons accelerated at
the FS and at the RS. Behind the shocks, we assume that particles
are accelerated to a power law distribution with $N(\gamma)\propto \gamma^{-p}$,
where $p$ is the power-law index of the electron
distribution\footnote{Recent particle-in-shell (PIC) simulations
  \citep{Spitkovsky:2008p7518} indicate that a pronounced Maxwellian
  electron component appears around the minimum of the non-thermal
  distribution in unmagnetized shocks. Observational consequences of
  this component are discussed in
  \citet{Giannios:2009p7522}. Furthermore, strong, large scale
  magnetic fields inhibit the formation of a non-thermal tail in the
  particle distribution, as found by the PIC simulations of
  \citet{Sironi:2009p414}. In Section 5, we comment on the
  observational implications of the lack of particle acceleration.}.

To compute the minimum Lorenz factor
$\gamma_{\rm min}$ of the electron distribution, we assume that a fraction
$\epsilon_e$ of the energy dissipated by the shock is used to
accelerate electrons. We use the same $\epsilon_e$ parameter for both
the RS and the FS. We estimate the maximum Lorenz factor $\gamma_{\rm
  max}$ to which electrons are accelerated by equating the synchrotron
cooling time with the gyration time. Then we have
\begin{equation}\label{eq:gmingmax}
  \gamma_{\rm min} = \dsfrac{3
    P_{\rm sh}}{\rho_{\rm sh} c^2}\dsfrac{m_p}{m_e}\dsfrac{p - 2}{p - 1} \epsilon_e\ ;\ 
  \gamma_{\rm max} = \left(\dsfrac{3m_e^2 c^4}{4\pi
      e^3B'}\right)^{1/2} 
\label{eq:gmin-gmax}
\end{equation}
where $P_{\rm sh}$ and $\rho_{\rm sh}$ are the thermal pressure and
comoving density of the fluid downstream of the shock, $m_e$ is the
electron mass, and $e$ is the electron charge.

Shock compression of the typical ordered external medium magnetic
fields is not sufficient to yield a substantial synchrotron emission
(see, however, \citealt{Kumar:11p7654} and Sect. 6 of this work for
possible exceptions) and a stochastic field $B'_{\rm st}$ amplified by
plasma instabilities in the shock is assumed. We model such a
stochastic field in a standard way, by assuming that its energy is a
fraction $\epsilon_B$ of the internal energy of the fluid behind the
FS, i.e. the comoving stochastic magnetic field strength is $B'_{\rm
  st}=\sqrt{24\pi \epsilon_B P_{\rm sh}}$. The same instabilities may
operate in the reverse shock (if the initial magnetic field of the
ejecta is weak). On the other hand, in high $\sigma_0$ models, the
shock compressed field dominates. Thereby, in order to have a
continuous description of the emission of magnetized and
non-magnetized models, for the purpose of computing the synchrotron
emissivity, we consider that the total comoving magnetic field
strength is the sum of the stochastic field plus the ordered
macroscopic field, i.e., $B'=B'_{\rm mac}+B'_{\rm st}$.

Using the SPEV code, we calculate the synchrotron emission from the
injected particles, and also follow the evolution of the particle
distribution downstream the shocks including cooling (adiabatic and
radiative) and synchrotron self-absorption
\citep{Mimica:2009p1445}. Our calculations neglect
synchrotron-self-Compton (SSC), which may make significant
contributions in the high-energy emission \citep{Fan:2008p7150} and,
in some cases, it can also affect the strength of the synchrotron
emission because of intensive Compton cooling of the electrons
\citep{Petropoulou:2009p7347}.

We note that SPEV takes a different approach from other methods such
as that of \citet{vanEerten:2010p7663}. While
\citet{vanEerten:2010p7663} perform a part of the emission calculation
during the RMHD simulation, SPEV operates on the post-processed data
of the RMHD simulation. To model the spatial distribution of
non-thermal particles it uses a volume-filling Lagrangian particle
scheme \citep[][Appendix A]{Mimica:2009p1445}. We modified SPEV so
that, if the synchrotron absorption is neglected, it computes the flux
density according to the equation 3 in \citet{Granot:1999mz} (see also
equation 4 in \citealt{Dermer:2008p2798}). In case when synchrotron
absorption is not neglected, we solve the radiative transfer equation
by following photon trajectories \citep[][Appendix A]{Mimica:2009p1445}.

\section{Dynamics and time scales}
\label{sec:dynamics}

The dynamics of the deceleration of the ejecta has been described in
detail in MGA09 for $\sigma_0=0$ and $\sigma_0=1$. Here, we have run
several additional simulations to investigate intermediate values of
$\sigma_0$. Since the methodology to compute the emission is much more
elaborated in the present paper (Sect.~\ref{sec:non-thermal}), we
have validated and refined the MGA09 findings (regarding the
observational signature of the models). The main results of MGA09 can
be summarized as follows:
\begin{itemize}

\item The deceleration depends on just two parameters $\xi$ and
$\sigma_0$. Lower $\xi$ and $\sigma_0$ result in stronger dissipation of energy
in the reverse shock and vice-versa.  

\item For the typical parameters of the GRB flows, $\sigma_0\sim 1$
  strongly suppresses the amount of energy dissipated in the reverse
  shock.  The front part of the ejecta initially forms a reverse
  rarefaction. At a later stage, because of the spherical expansion, a
  weak reverse shock forms at the rarefaction tail.

\item The total magnetic energy of the ejecta remains approximately 
constant or even increases while they are crossed by the reverse shock 
(also discussed in \citealt{Zhang:2005ts}).

\item After the RS reaches the rear end of the ejecta, a rarefaction 
forms and crosses the ejecta and the shocked external medium within
the expansion timescale of the blast wave. The rarefaction dilutes the magnetic energy 
in the ejecta. When the rarefaction head catches up with the FS there is a
moderate but sharp decrease of the Lorenz factor of the shock. 
Only at this stage the solution approaches
the self similar evolution of \citet{Blandford:1976yg}.
 
\end{itemize}

A common feature of all our simulations is that the self-similar
solution is established at an observer time that is longer than the
duration of the burst (which corresponds to the light-crossing time of
the shell thickness).  At earlier times, there is substantial energy
injection at the FS that clearly affects the rate of
deceleration of such shock (see Sect.~\ref{sec:lightcurves} for
lightcurves and Sect.~7 for observational implications of this
effect).  The self-similar evolution can, therefore, be assumed only
when calculating afterglow lightcurves well after the duration of the
burst.

While we cannot exclude that there exists a region of the ($\xi
-\sigma_0$) parameter space in which the energy transfer takes place
before the end of the GRB, we consider this possibility unlikely. The
reasoning is the following. It can be shown that the observer time
corresponding to the reverse shock crossing of the ejecta is larger
than the duration of the burst \citep[e.g.,][]{Nakar:2004p7481};
magnetization of the ejecta does not change this conclusion 
\citep[e.g.,][]{Fan:2004if}. At this stage a rarefaction forms and crosses the
shocked ejecta. Our simulations show that energy is transferred at the
FS until the rarefaction reaches the FS. Since the rarefaction moves
slower than light, the energy transfer is completed after the peak of
the reverse shock emission (which trails the bulk of the GRB
emission).

\subsection{The long-term evolution of the magnetic energy}
\label{sec:longterm}

Although, as we just discussed, the bulk of the energy of the ejecta
is passed onto the shocked ISM on a short time scale (slightly longer
than the GRB duration), there is a residual magnetic energy remaining
in the ejecta.  Such energy may, however, have important observational
implications if dissipated at later times. For example, it is argued
in \citet{Giannios:2006hb} that the slowing down of magnetized ejecta
leads to revival of (current driven) MHD instabilities.  The
instabilities can lead to the dissipation of magnetic energy in
localized reconnection regions which may power the observed afterglow
X-ray flaring \citep{Burrows:2005kk}. In this picture no late central
engine activity is needed. Our 1D models cannot tackle the issue of
the nature of such (3D) instabilities, but they can quantify how much
energy remains in the magnetized shell as function of observer's time
that could potentially power any afterglow flaring.

\begin{figure}
\includegraphics[width=8.5cm]{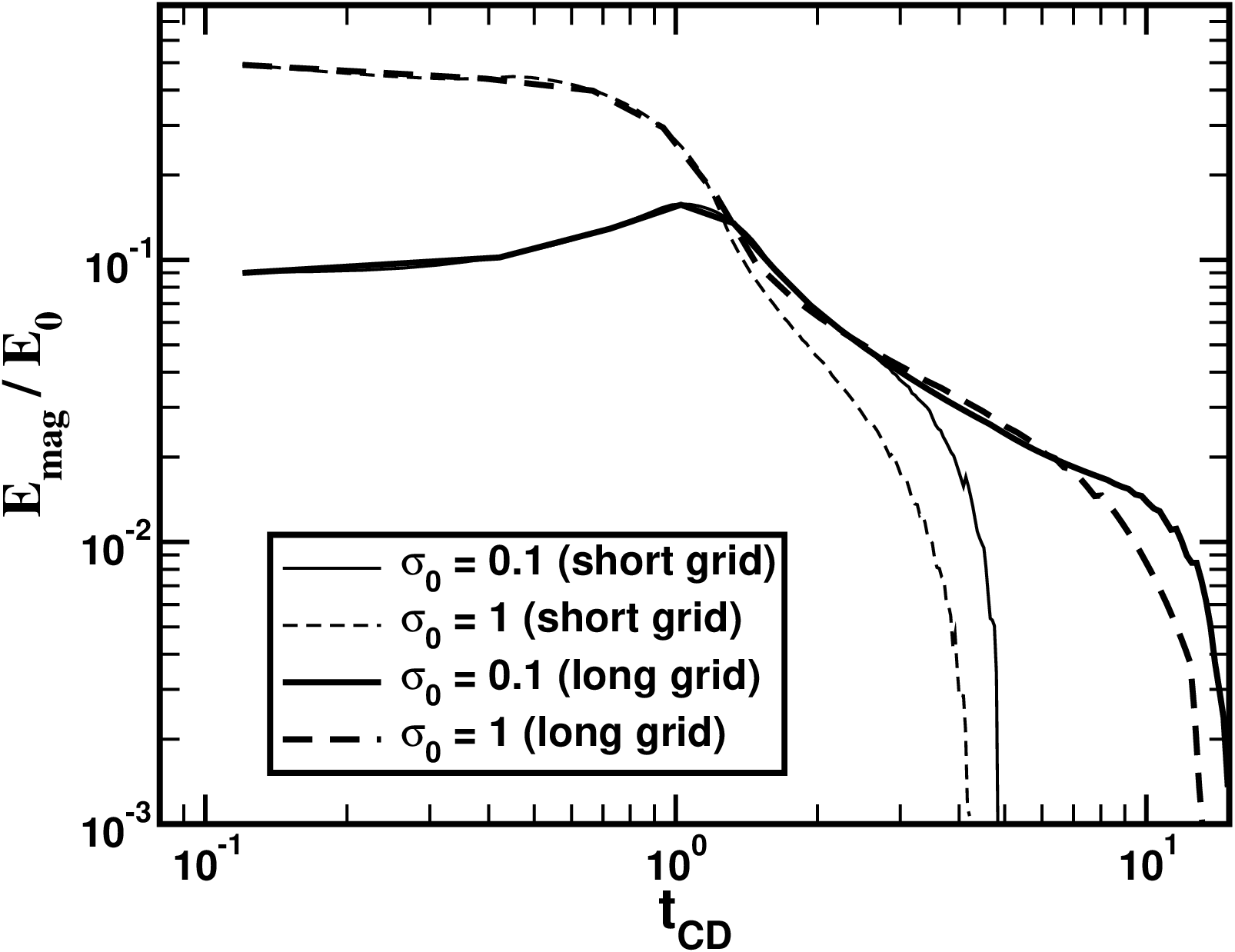}
\caption{ Magnetic energy fraction as 
  function of the dimensionless observer time $t_{\rm CD}$. Thin lines
  show results for thin shell models $\sigma_0 = 0.1$ and $\sigma_0 =
  1$ with the short grid, while the thick lines show results for the
  long grid. The rapid drop of the magnetic energy at late times is
  not physical but due to the fact that the whole magnetized shell
  leaves our computational box. This problem is ameliorated in the
  long grid models, where it shows up at later times, when the
  magnetic energy content is much smaller. }
\label{fig:e_mag}
\end{figure}

In \figref{fig:e_mag}, we plot the magnetic energy as function of a
dimensionless observer time defined as $t_{\rm CD}=(ct-R_{\rm
  CD})/\Delta_0$, where $R_{\rm CD}$ is the location of the contact
discontinuity separating the shocked circumburst medium from the
ejecta.  Since any emission from magnetic dissipation takes place
behind the contact, $t_{\rm CD}$ gives the shortest time after the
burst that any emission from the ejecta will be observed. 

The remapping technique we are using to track the evolution of the
shell consists in adding a certain number of numerical zones in front
of the leading forward shock. The same number of zones is removed from
the rear part of the domain. Since the shell is ultrarelativistic, the
rear boundary is causally disconnected from the RS during the time in
which we evolve our models. Thus, even considering that a part of the
shell is lost through the rear radial boundary, for the purpose of
computing the dynamics (or the emission) of the leading forward and
reverse shocks, it suffices a relatively small computational domain
$L$ spanning not only the shell, but a small portion of the leading and
of the trailing medium, namely $L=1.5\Delta_0$ ($L=4\Delta_0$) for the
thick (thin) shell models. However, due to the loss of the rear part
of the shell, the calculation of any integral quantity extending over
the whole shell (in particular the magnetic energy) is
underestimated. To ameliorate such a deficiency of our models, we have
computed some of the "thin shell" simulations with a $\sim 3$ times
larger grid (``long grid'' in Fig.~\ref{fig:e_mag}) than that used in
MGA09 (``short grid'' in Fig.~\ref{fig:e_mag}).  At $t_{\rm CD}<1$ the
magnetic energy changes little (and actually increases in the
$\sigma_0=0.1$ run because of the (reverse) shock compressing the
ejecta). The magnetic energy drops fast at $t_{\rm CD}\sim 1$ because
of the rarefaction that follows.  Nevertheless, a fraction of $\sim
10$\% of the total energy remains in magnetic form for longer time
showing only modest decline with time (note that the large grid is
essential to follow this late evolution). We find that at least $\sim
2\%$ of the energy in still in magnetic form at $t_{\rm CD} \sim 10$
(i.e., about 10 times the burst duration). The rapid drop of the
magnetic energy at late times is not physical. It is due to the fact
that the contact discontinuity reaches the back of our grid and the
whole magnetized shell leaves our computational box (and we are left
with the shocked ISM).

In view of these results, we suggest that there is sufficient energy
remaining in the magnetic field for at least $\sim 10$ times the
duration of the burst to yield a substantial emission if a large part
of the magnetic energy is dissipated and radiated away.

\section{Light curves and spectra}
\label{sec:lightcurves}

In this section we present the non-thermal light curves and spectra
computed from our models. We first discuss the generic light curves
corresponding to thin and thick shells for different magnetization of
the ejecta. We also show a prototype emission spectrum at the end
of this section.

\subsection{Thin shells}
\begin{figure}
\includegraphics[width=8.5cm]{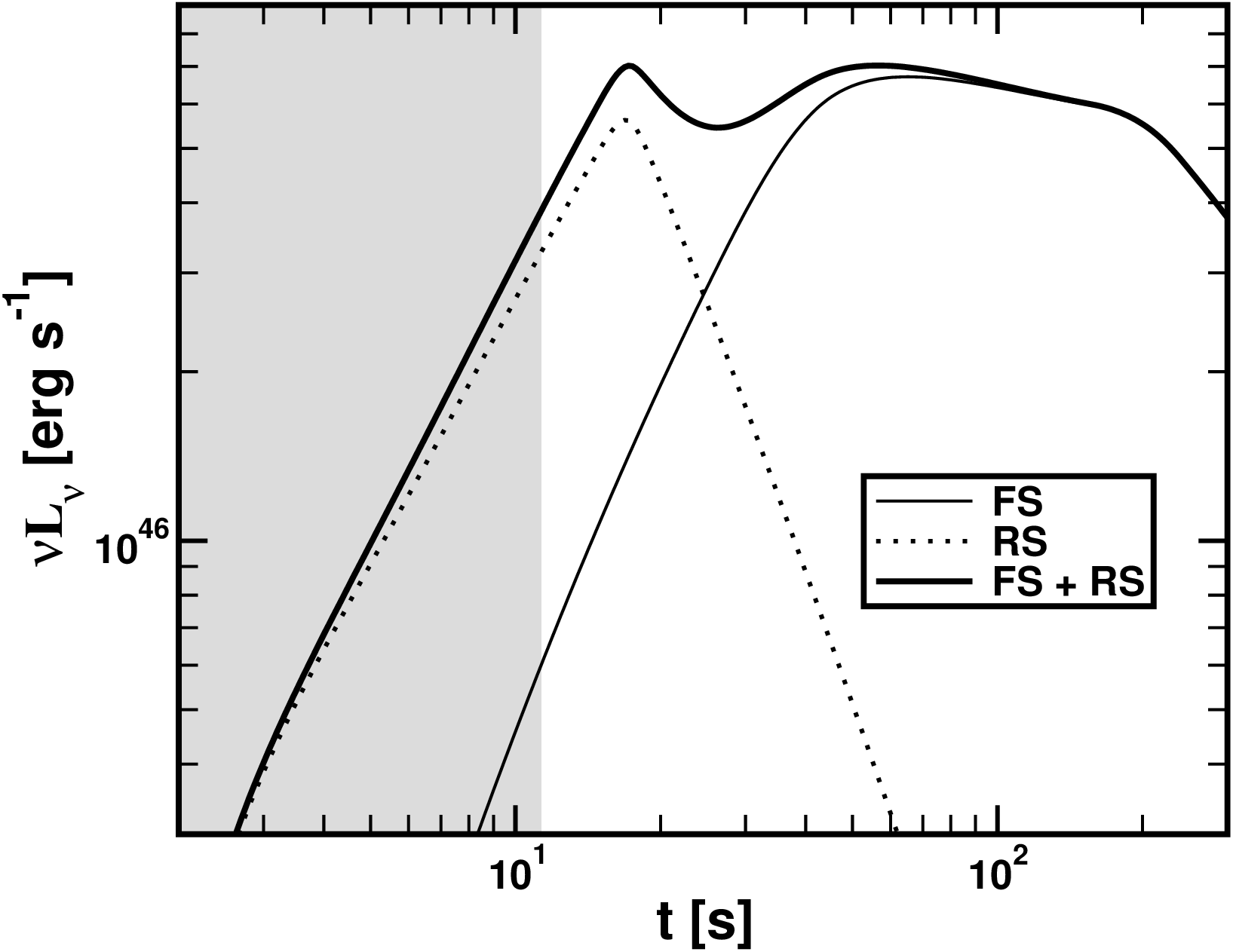}
\caption{R-band light curve for the thin shell with initial
  magnetization $\sigma_0=0.01$. Shown are the light curves originating
  from the FS (thin full line), RS (dotted line) and the combined
  light curve (thick full line). In the calculation $\epsilon_B =
  5\times 10^{-3}$, $\epsilon_e = 10^{-1}$ and the electron power-law
  index $p = 2.25$. The shaded area denotes the observational interval
  which is simultaneous with the prompt emission.}
\label{fig:FSRS-thin}
\end{figure}

\begin{figure}
\includegraphics[width=8.5cm]{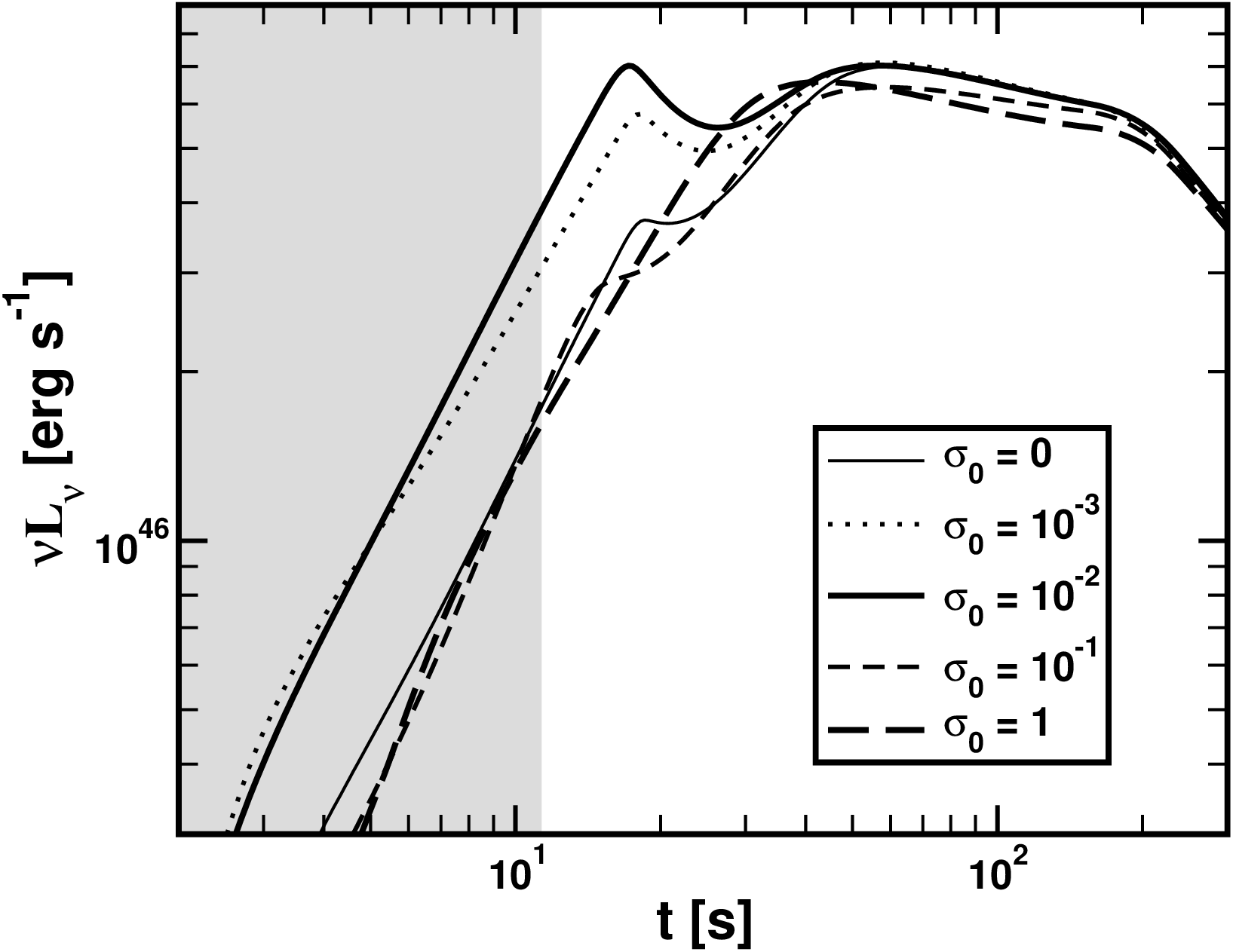}
\caption{Combined (FS + RS) R-band light curves for thin shell models
  with $\sigma_0 = 0,\ 10^{-3},\ 10^{-2}, \ 10^{-1}$ and $1$ (thin
  full, dotted, thick full, thin dashed and thick dashed lines,
  respectively). The rest of the parameters are like in
  \figref{fig:FSRS-thin}.}
\label{fig:R-band-thin}
\end{figure}

Figure~\ref{fig:FSRS-thin} shows the R-band light curves for the thin
shell (i.e., $\xi=1.1$) model with initial magnetization $\sigma_0=
0.01$ and bulk Lorenz $\Gamma_0=300$. The emission from the shocked
external medium (thin full line), shocked ejecta (dotted line) and the
total emission (thick full line) is displayed. We have assumed
$\epsilon_B = 5\times 10^{-3}$ and $\epsilon_e = 10^{-1}$ so as to
calculate emission. The total R-band light curves for models with
different values $\sigma_0$ from $0$ to $1$ are shown in
\figref{fig:R-band-thin}.

As can be seen in \figref{fig:FSRS-thin} and, in general, in
\figref{fig:R-band-thin}, for the parameters chosen the RS emission is
marginally brighter than that of the FS in optical bands.  Moderate
differences in light curves for $t\simless 25\,$s are caused by the
effect of $\sigma_0$ on the RS emission.  Increasing $\sigma_0$ from
$0$ to $0.01$ leads to progressively brighter RS emission because of
increased synchrotron emissivity resulting from a stronger
field. Further increase of the magnetization greatly weakens the RS
emission and the amount of heating that takes place at the shock
(and, therefore, reduces the electron minimum Lorenz factor in the
downstream fluid; see Eq.~\ref{eq:gmin-gmax}) resulting in a
negligible optical emission compared to that of the FS.

\begin{figure}
\includegraphics[width=8.5cm]{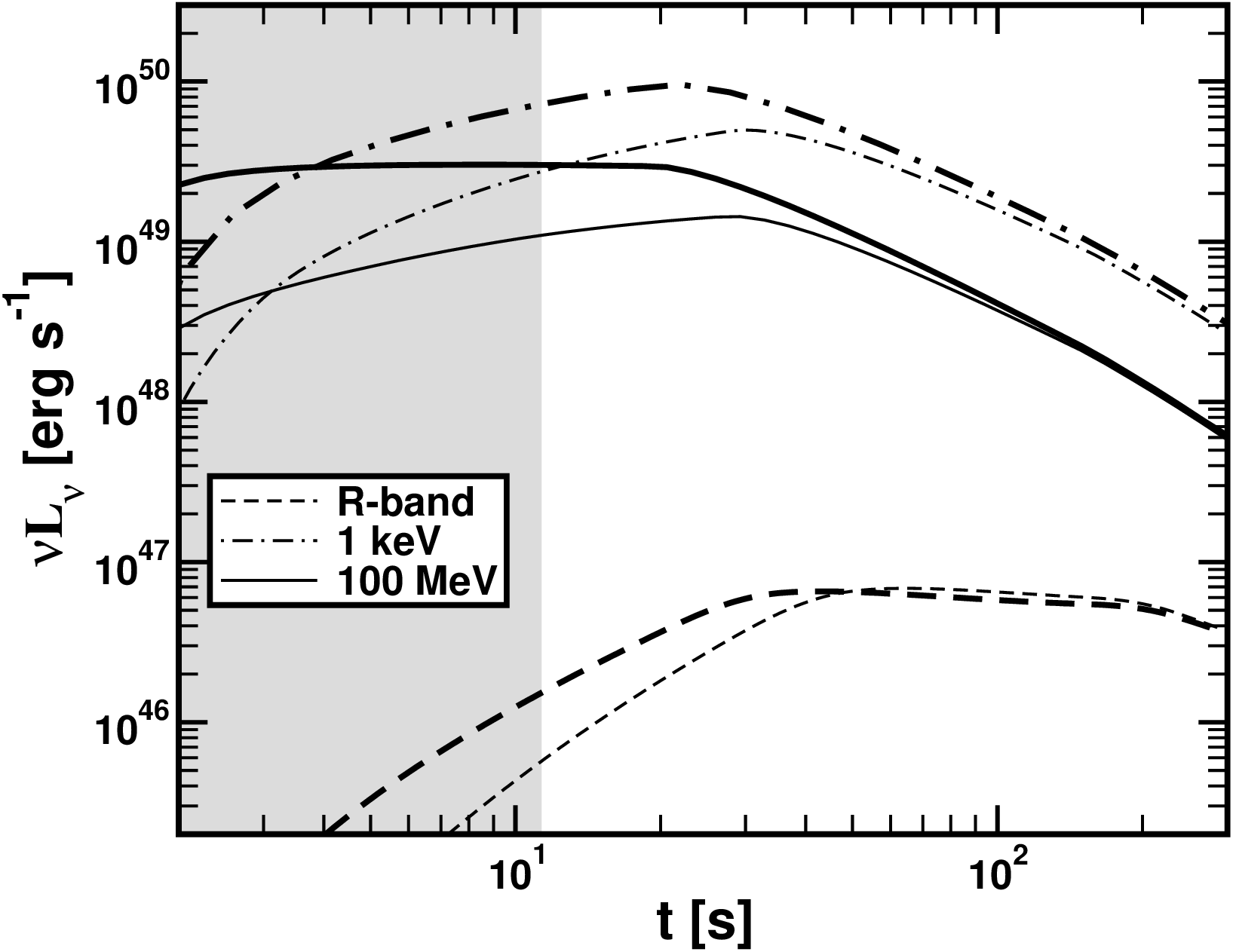}
\caption{Forward shock emission. R-band, $1\,$keV and $100\,$MeV light
  curves (full, dotted and dashed lines, respectively) for thin shell
  models with $\sigma_0 = 0$ (thin lines) and $\sigma_0 = 1$ (thick
  lines).}
\label{fig:RXG-thin}
\end{figure}

The FS emission shows a weak dependence on $\sigma_0$, the differences
being more pronounced at higher energies (X-rays, gamma-rays) as shown
in \figref{fig:RXG-thin}. Our models display an achromatic break in
the light curves shortly after the end of the burst. The break takes
place at $t\sim 30\,$s and $20\,$s for the $\sigma_0=0$ and
$\sigma_0=1$ models, respectively, and it results from the rarefaction
that reaches the FS and decelerates it. At early times, the emission
in gamma- and X-rays is systematically brighter and flatter for
high-$\sigma_0$ models than in the optical band, while the gamma- and
X-ray flux rises faster for low-$\sigma_0$ models. Observing
early enough after the burst, the high-energy emission from the FS can
contain a valuable information about the magnetization of the
ejecta. If the afterglow is observed too late, in this example after
approximately $30\,$s(\figref{fig:RXG-thin}), light curves with
radically different values of $\sigma_0$ become indistinguishable,
since the ejecta enters into the self-similar evolution phase,
dependent only on the total initial energy and the external medium
density, which are the same for all models.

\subsection{Thick shells}

\begin{figure}
\includegraphics[width=8.5cm]{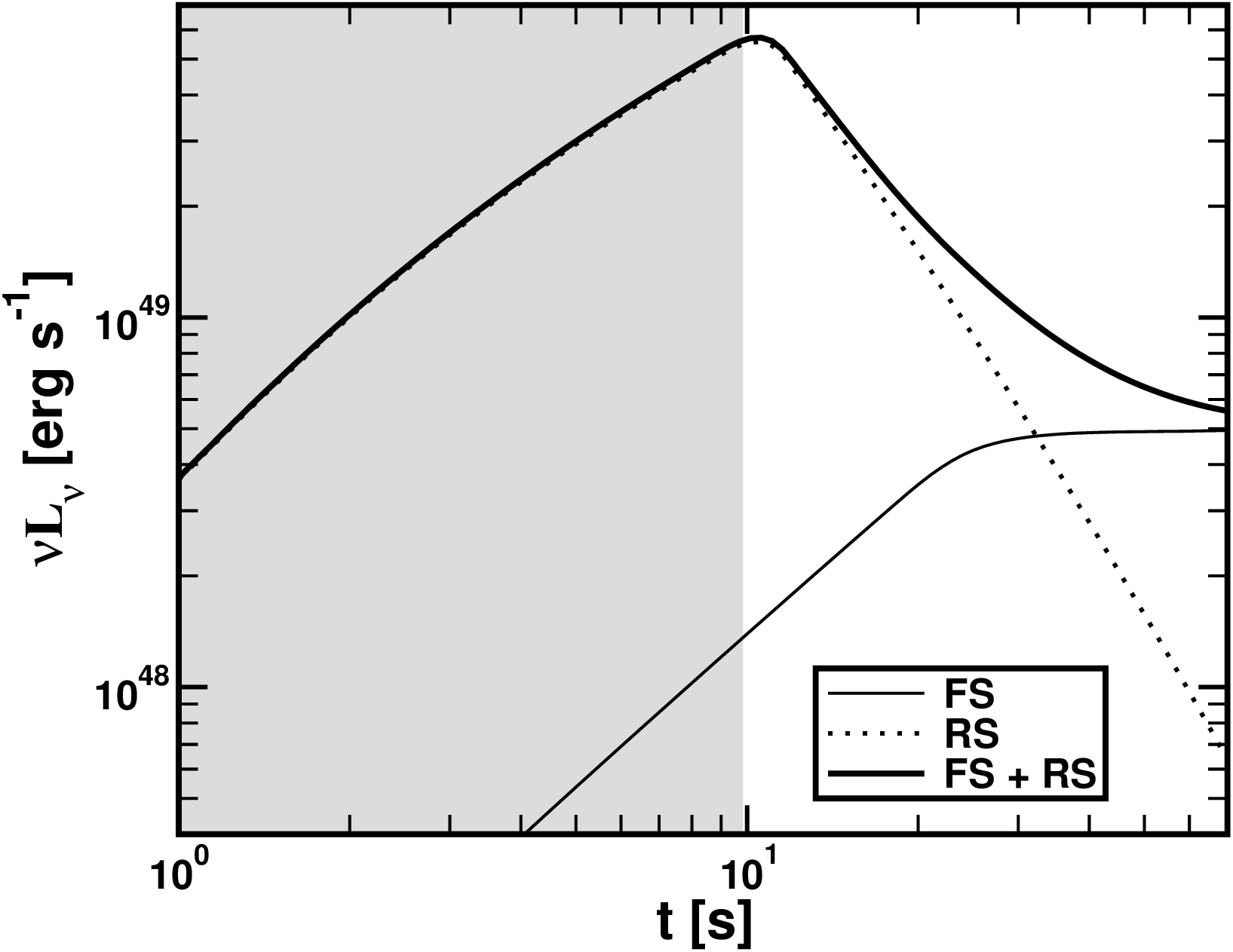}
\caption{Same as Fig.~\ref{fig:FSRS-thin} but for the thick shell with initial
  magnetization $\sigma_0=0.01$. }
\label{fig:FSRS-thick}
\end{figure}

We, now turn to the emission from a thick shell ($\xi=0.5$) for
$\Gamma_0=750$. Figure~\ref{fig:FSRS-thick} shows the contributions
from the FS and the RS to the total R-band light curve for the model
with initial magnetization $\sigma_0 = 0.01$. We can see that the RS
the contribution to the emitted flux here is much more important than
in the thin shell case. The RS dominates the emission until $t\simeq
50\,$s.

\begin{figure}
\includegraphics[width=8.5cm]{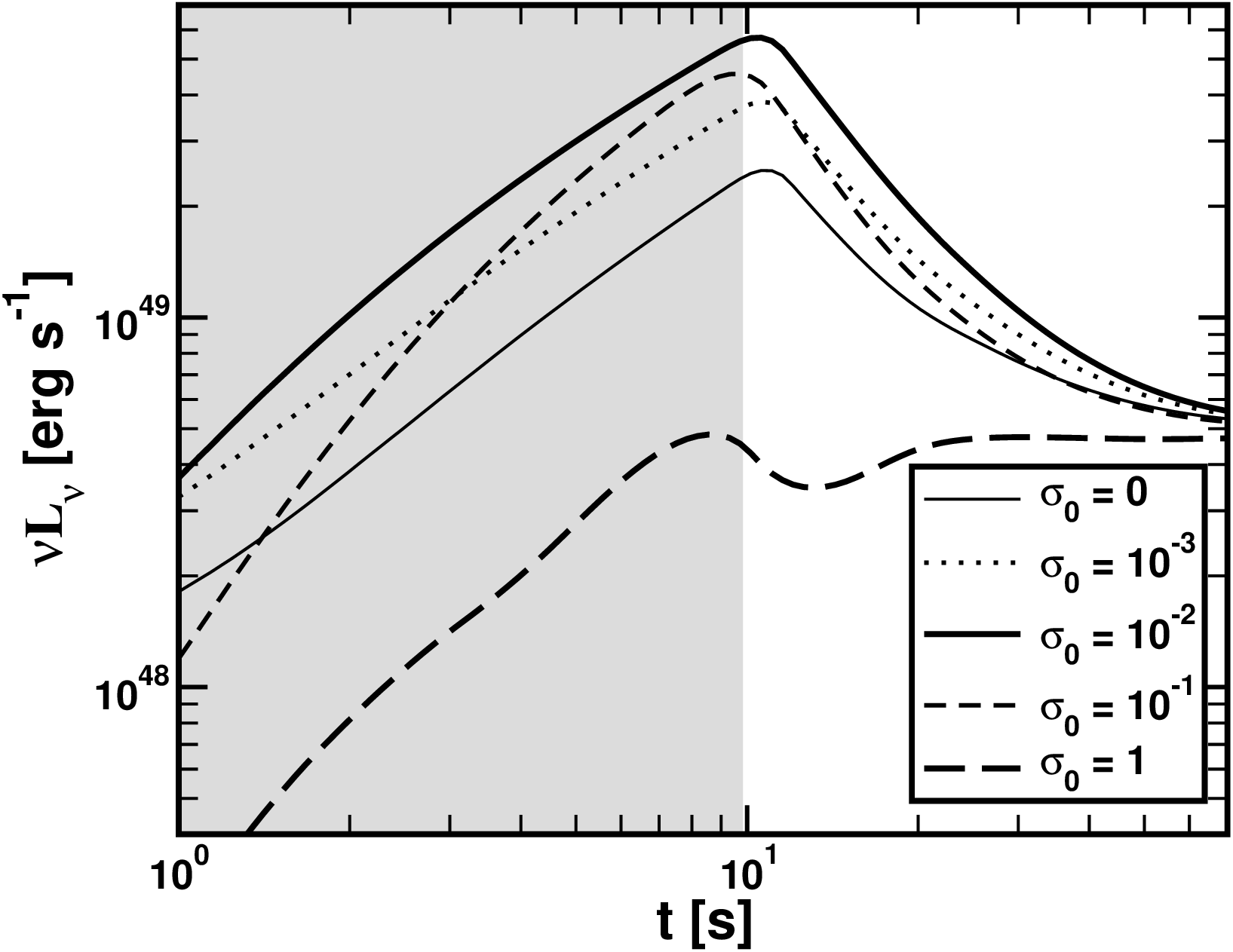}
\caption{Same as Fig.~\ref{fig:R-band-thin}, but for thick shell models.}
\label{fig:R-band-thick}
\end{figure}

As can be seen in \figref{fig:R-band-thick}, the RS emission dominates
in the early afterglow over that of the FS in the optical bands.
Large differences in light curves for $t\simless 50$ s are caused by
the effect of the magnetization $\sigma_0$ on the RS emission. For
$\sigma_0=0$, there is a clear RS signature.  It results from the
synchrotron emission of non-thermal particles gyrating in the
stochastic magnetic field (presumably) generated at the shocks.
Increasing $\sigma_0$ from $0$ to $0.01-0.1$ leads to brighter RS
(synchrotron) emission because of the stronger (shock compressed)
magnetic field. Maximal RS emission happens for $\sigma_0\sim
0.01-0.1$. Around such values of $\sigma_0$ there is a turnover of the
emission, which drops as the magnetization keeps growing. In effective
terms, the RS light curve changes only moderately for
$10^{-3}\simless\sigma_0\simless 0.1$, and hence, inferring an accurate
value of $\sigma_0$ from a careful modeling of the RS emission is
difficult. Further increase of the magnetization greatly weakens the
reverse shock and suppresses the energy injected to the particles. As a
result, there is a clear suppression of the reverse shock emission for
$\sigma_0 \sim 1$ or higher.

\begin{figure}
\includegraphics[width=8.5cm]{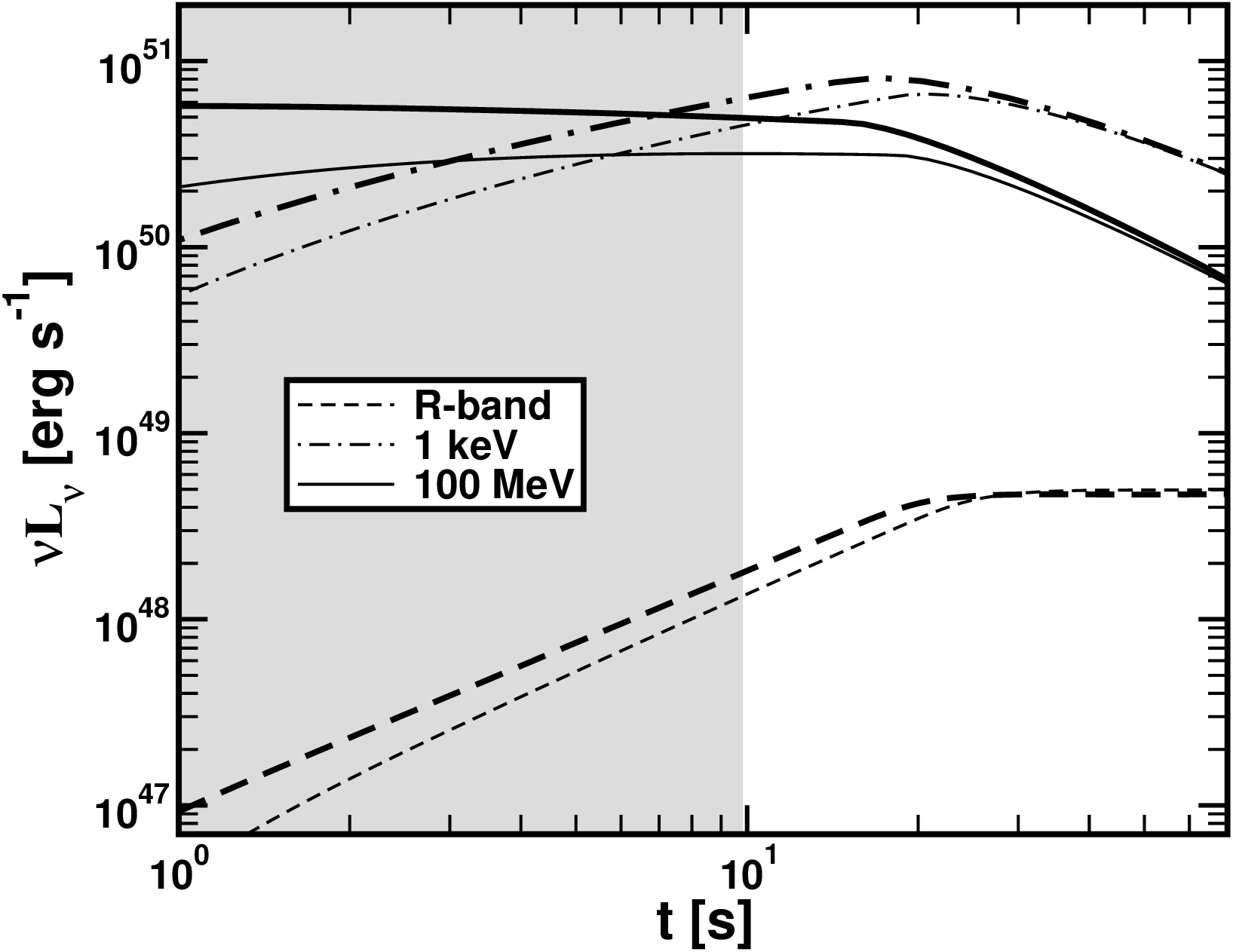}
\caption{Same as \figref{fig:RXG-thin}, but for thick shell models.}
\label{fig:RXG-thick}
\end{figure}

Other mild differences in the optical lightcurves are caused by
differences in the Lorenz factor of the FS at early times because of
the different $\sigma_0$ as shown in \figref{fig:RXG-thick}. The thick-shell models 
show the same achromatic break in the light curves shortly after the
end of the burst as seen in the thin shell ones.  In this example, after
approximately $30\,$s, FS light curves become indistinguishable, since
the ejecta evolution approaches to the self-similar evolution phase.

\subsection{Emission spectra}

\begin{figure}
\includegraphics[width=8.5cm]{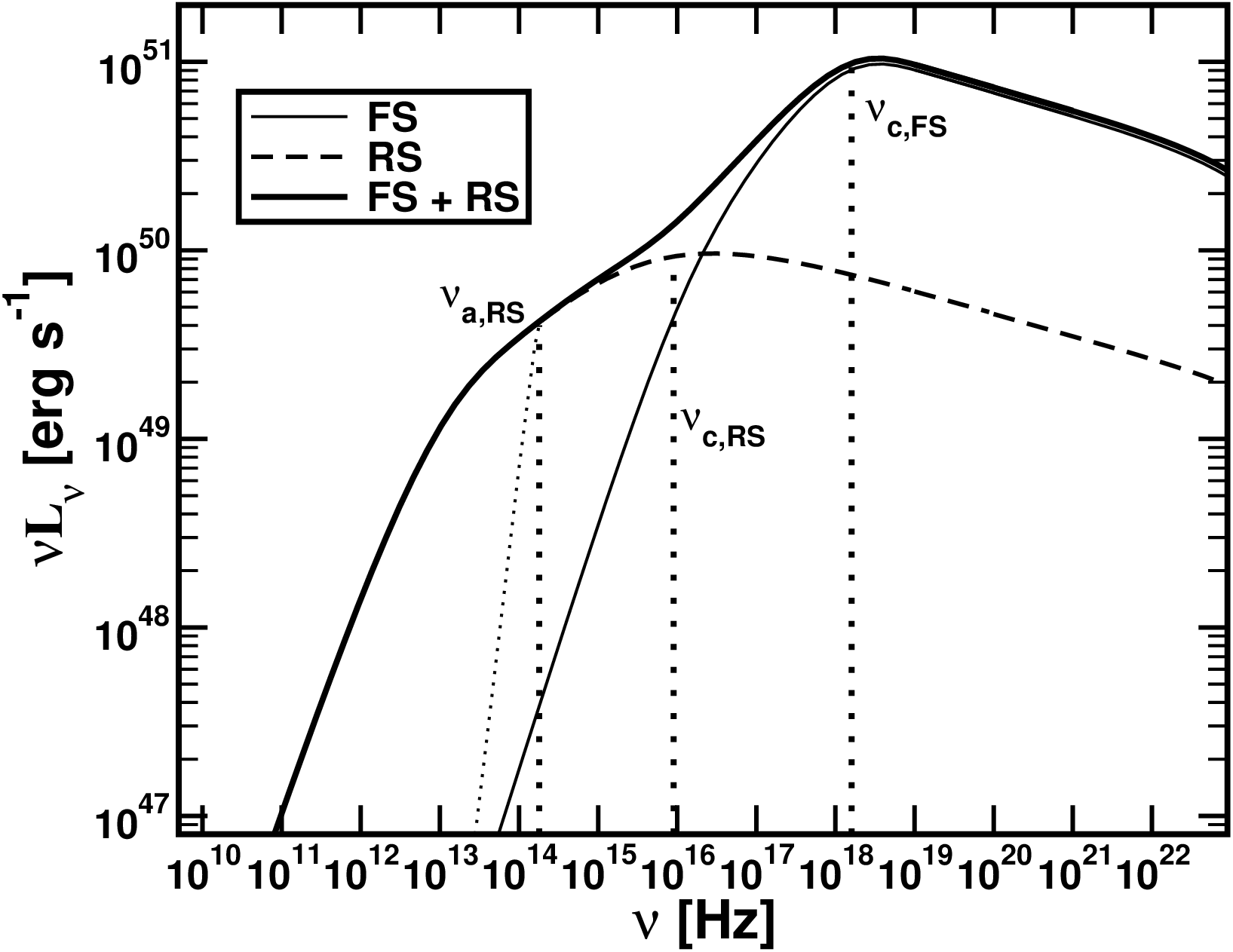}
\caption{Emission spectrum at $t = 10\,$s for the thick shell model
  with $\sigma_0 = 0.01$. Thin, full and dashed lines show optically thin FS
  and RS spectra, respectively. Thick full line shows the total
  spectrum. Thin dotted line shows the portion of the RS spectrum
  which is optically thick to synchrotron self-absorption (see text for
  details). Vertical dotted lines denote
  FS and RS cooling break frequencies $\nu_{\rm c,FS}$ and $\nu_{\rm c,RS}$,
  as well as the RS synchrotron-self absorption break frequency $\nu_{\rm
    a,RS}$.}
\label{fig:spec-thick}
\end{figure}

The instantaneous spectrum of the thick shell model $\sigma_0=0.01$
taken at $t=10\,$s is shown in \figref{fig:spec-thick}. This time
approximately corresponds to the peak of the RS emission and to the
end of the prompt ($\sim$MeV) emission.
Thick full line shows the total spectrum, while thin full and dashed
lines show the FS and RS contributions, respectively. Both spectra
have been computed neglecting the synchrotron self-absorption
(SSA). Thin dotted line shows the RS spectrum when SSA is taken into
account\footnote{We do not include SSA in all the light curve
  calculations since the computational costs of such a calculation are
  prohibitive (one of the main reasons is that the algorithm which
  includes SSA is not as easily parallelizable as the optically-thin
  calculation). However, we have verified that the light curves of all
  the bands shown in this work lie above the SSA frequency $\nu_{\rm
    a,RS}$, making the SSA correction negligible for the purpose of
  the present paper.}.

The peak of the $\nu L_{\nu}$ emission produced by each of the shocks
corresponds to their respective cooling break frequencies. It appears
in the X-ray band and in the UV band for the FS and RS, respectively.
In \figref{fig:spec-thick} we also show the frequency of the SSA break
for the reverse shock emission ($\nu_{\rm a,RS}$) that lies in the
infra-red. Not surprisingly, the optical emission in the range $\sim
10^{14}-10^{15}\,$Hz is dominated by the RS, while the
X-ray-to-$\gamma$-ray emission is dominated by the FS.

In deriving these results, we have assumed that electrons are
accelerated into a power law distribution at the shocks.  Despite the
fact that particle acceleration in (relativistic) shocks is poorly
understood, there are indications that strong, large scale fields
along the shock plane can inhibit particle acceleration. For instance
recent PIC simulations by \citet{Sironi:2009p414} have shown that for
$\sigma\simmore 0.01$, magnetized, relativistic shocks result in
efficient heating of the particles with the distribution lacking
non-thermal tails. If this result holds for the mildly relativistic
shock forming in the ejecta, such lack on high-energy tail can
substantially affect the RS emission (and, as a matter of fact, any
emission from internal shocks). If the characteristic frequency
(associated to the thermal distribution of the particles) lies bellow
the observed band then there will be practically no observed emission
from the RS. When the characteristic frequency of the RS emission (as
is often the case) lies in the infrared, bursts of sufficiently high
$\sigma_0\simmore 0.01$ will be optically dim (there are no electrons
accelerated to high enough energy to (synchrotron) emit in the optical
or harder bands). This is an additional potential mechanism that
suppresses the RS emission in moderately to strongly magnetized
ejecta.

\section{Application to GRB 990123 and GRB 090102}
\label{sec:application}

In this section we apply our early afterglow models to two specific
GRBs exhibiting bright optical emission which may be powered by a
reverse shock, GRB~990123 \citep{Briggs:1999p7488,Akerlof:1999yf} and
GRB~090102 \citep{Gendre:2009p6929,Steele:2009p5925}. The famous
GRB~990123 had a 9th magnitude optical flare a few seconds after the
end of the GRB and is one of the brightest ever detected.  090102 is a
rather bright GRB whose early optical afterglow shows declining, but
bright emission \citep{Gendre:2009p6929} and, most importantly, a
significant $\sim 10\%$ polarization, indicating large scale fields in
the emitting region \citep{Steele:2009p5925}.

Several of the parameters of our model can be directly inferred by the
observed properties of these bursts. The rest-frame duration of both
bursts is $T_{\rm GRB}\simeq 10\,$s, constraining the shell thickness
to be $\Delta_0\simeq cT_{\rm GRB}\simeq 3\times 10^{11}\,$cm. The
gamma-ray fluence of GRB~990123 and 090102 is $E_\gamma \simeq 3\times
10^{54}\,$erg and $E_\gamma \simeq 6\times 10^{53}\,$erg,
respectively. Assuming $\sim 20\%$ efficiency for the prompt emission
mechanism the total (isotropic equivalent) energy of the ejecta can be
estimated to be $E_{990123} \simeq 1.5\times 10^{55}$ erg and
$E_{090102} \simeq 3\times 10^{54}$ erg. Finally, the fact the the
reverse shock peaks very shortly after the burst indicates that
$\xi\simless 1$ in these bursts.

\begin{figure}
\includegraphics[width=8.5cm]{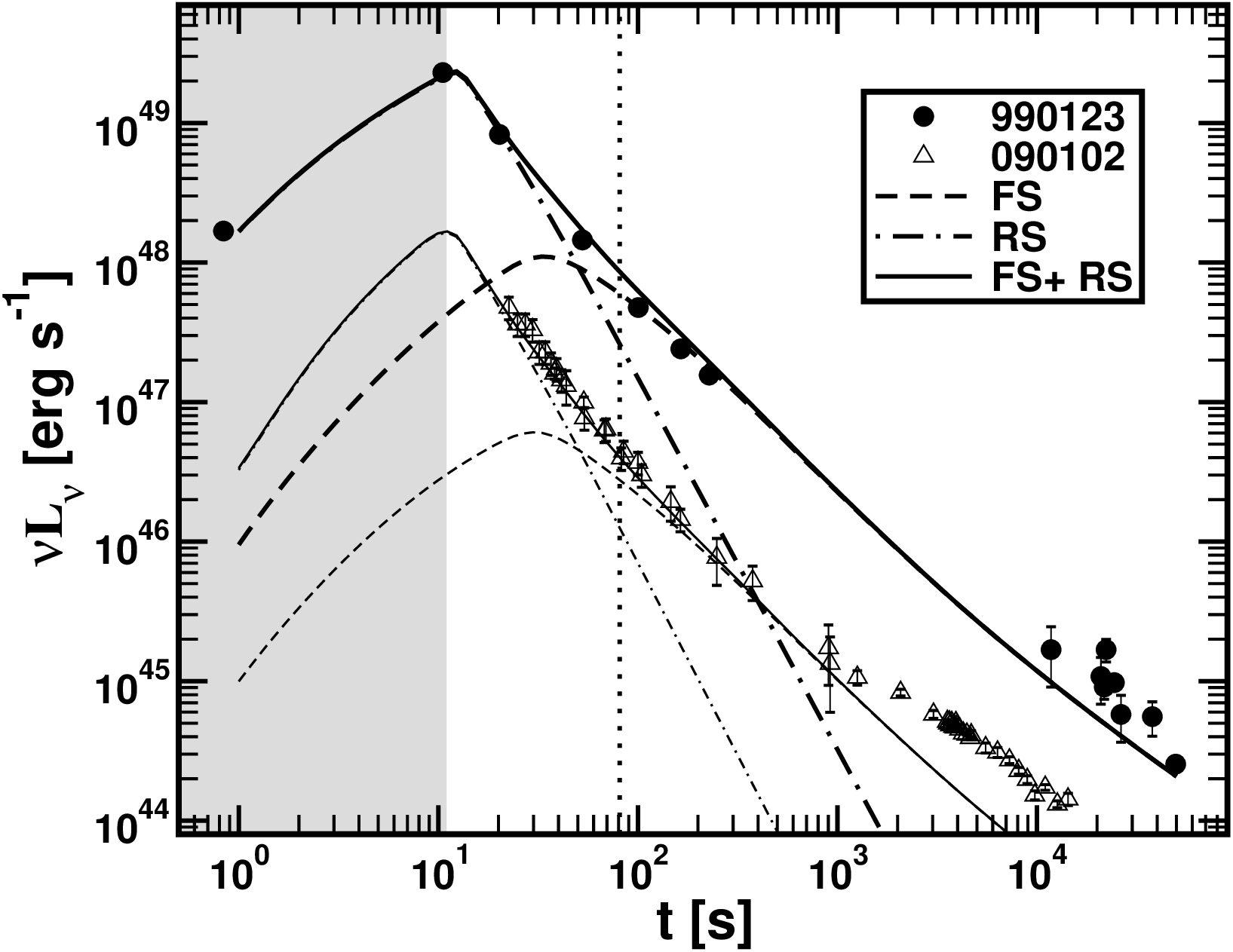}
\caption{R-band light curve for GRB 990123
  \citep{Briggs:1999p7488,Akerlof:1999yf}, GRB 090102
  \citep{Gendre:2009p6929,Steele:2009p5925} plotted in the rest frame
  of the central engine of the burst. Shown are the thick shell
  models which most closely match the observed light curves, assuming
  for both bursts ejecta width $\Delta_o=3.3\times 10^{11}\,$cm 
  (with corresponding duration of $11$ seconds; marked by the shaded
  region in the plot).  The $\sigma_0 = 0.01$ model (thick lines) has
  been rescaled to $\Gamma_0 = 640$ and $n_{\rm ext}=10\,$cm$^{-3}$. 
The $\sigma_0 = 0.1$ model (thin lines) has been rescaled to 
$\Gamma_0 = 940$, $n_{\rm ext}=1$\,cm$^{-3}$. For both curves 
$\epsilon_e = 0.02$ and $\epsilon_B = 4\times 10^{-7}$ are assumed. Full, dot-dashed and dashed lines
  denote the total (FS + RS), RS and, FS emission. The vertical dotted
  line shows the observer time up to which the data from the
  simulation has been used to compute the FS emission; from that point
  onwards the Blandford-McKee asymptotic solution has been used to
  compute the FS light curve.}
\label{fig:R-band-app}
\end{figure}

Taking these observational constraints into account, we look for the
values $\Gamma_0$ and $\sigma_0$ for which our rescaled numerical
models best fit the observations.  Figure~\ref{fig:R-band-app} shows
the R-band light curves of GRB~990123 (circles) and GRB~090102
(triangles)\footnote{To convert flux to luminosity, we have assumed
  for both GRBs a redshift $z = 1.6$, and a standard cosmology model
  with the Hubble and density parameters are $H_0 =
  71~\,$km\,s$^{-1}\,$Mpc$^{-1}$, $\Omega_m = 0.27$, and
  $\Omega_\Lambda = 0.73$.}. We also plot the light curves of two
thick shell models, the $\sigma_0 = 0.01$ rescaled to $\Gamma_0 = 640$
(thick lines) and $\sigma_0 = 0.1$ rescaled to $\Gamma_0 = 940$ (thin
lines). Full, dot-dashed and dashed lines denote the total, RS, and FS
emission. The data in our simulation only allowed us to compute the FS
emission up to $t \simeq 80\,$s. By this time, however, the evolution
of the blast is self-similar.  For the purpose of better
visualization, we have used the Blandford-McKee asymptotic solution to
extend the FS light curve to later times.

Even though our best fits suggest that $\sigma_0 = 0.1$ and $\sigma_0
= 0.01$ for GRBs 090102 and 990123 respectively, $\sigma_0$ is not
well constrained. From \figref{fig:R-band-thick}, one can see that the
RS emission varies relatively little for a large range of $0.01
\simless \sigma_0 \simless 0.1$ indicating that pinpointing the
magnetization with precision from fitting the reverse shock light curve
is hard.

The optical afterglow of GRB~090102 shows $\sim 10$\% polarization at
an exposure that took place $160-220\,$s after the GRB trigger
\citep{Steele:2009p5925}. This value is very hard to understand as
synchrotron emission from small scale magnetic fields but is still
smaller than that expected from a coherent field.  The exposure time
interval of the \citet{Steele:2009p5925} observations corresponds to
the time interval $100-140\,$s in the frame of the source. One can see in
\figref{fig:R-band-app} that at $\sim 100\,$s the FS emission may
already dominate over the RS emission by a modest factor. The observed
10\% polarization may, therefore, be understood as a combination of the very
weekly polarized, brighter FS emission and a strongly polarized emission
coming from RS that propagates in a flow of coherent field. If this interpretation is
correct, earlier polarimetry of afterglow emission should reveal even
stronger polarization signal, for these few bursts for which the RS emission is
sufficiently bright with respect to the FS.

Another intriguing result from the fitting of our models is that, in
order for the FS to account for the late-time emission, a rather small
value of $\epsilon_B\sim 10^{-7}$ needs to be assumed. For GRB~990123,
a similar finding is presented in \citet{Panaitescu:2004p7344}. Note
that such weak fields may be simply due to shock compression of the
ISM magnetic fields (i.e., there is no need for field generation at the
shocks for these bursts). \citet{Kumar:11p7654} present more bursts
whose afterglow modeling is compatible with very low values of
$\epsilon_B$.

\section{Discussion and conclusion}
\label{sec:discussion}

The theoretical modeling of GRB central engines is still far from
predicting whether the GRB flow is thermally or magnetically dominated
in the launching region. Furthermore, prompt GRB emission mechanisms
are not understood well enough to enable probing of the magnetization
of the emitting plasma.  A more promising way to infer the
magnetization of the GRB flow may come through studying interactions
with the circumburst medium. Quantitative inferences, however,
critically depend upon the detailed modeling of the dynamics of such
interaction, as well as the emission mechanisms from shocked
plasma. In this work we presented relativistic MHD models of
ejecta-medium interactions coupled to a radiative transfer code to
calculate multi-wavelength synchrotron lightcurves. The most salient
results of our parametric study are the following:

\begin{enumerate}

\item For very weakly magnetized ejecta, the standard forward/reverse
  shock structure \citep{Sari:1995oq} is reproduced by our
  simulations. The reverse shock can dominate the early optical/IR
  emission while the forward shock dominates the subsequent emission
  at all wavelengths.  Increasing the magnetization of the ejecta to
  $\sigma_0\sim 0.01-0.1$, the reverse shock remains strong while the
  powerful fields result in a brighter synchrotron emission,
  typically, in the optical band. At still stronger magnetization
  $\sigma_0\simmore 1$, the dissipation at the reverse shock and,
  therefore its emission are greatly reduced.

\item Observationally, the majority of GRBs do not show bright optical
  emission at early times. \citet{Klotz:2009p4221} present a
  systematic study of early optical observations and tight upper
  limits tens of seconds after GRB emission. We interpret this lack of
  early optical detections as evidence for weak/absent reverse shock
  as expected from GRB flows characterized by $\sigma_0\simmore 1$ at
  large distance from the central engine. Nevertheless, a few GRBs
  such as 990123 and 090102 do show bright optical emission that is
  likely coming from the reverse shock
  \citep{Akerlof:1999yf,Meszaros:1999p4433,Nakar:2005p4038,Steele:2009p5925}.

\item Our detailed modeling of early afterglow emission from
  GRB~990123 and 090102 indicates that the magnetization of the ejecta
  is $\sigma_0\sim 0.01-0.1$ for these bursts, i.e., close to the
  value that maximizes the RS brightness (see
  \figref{fig:R-band-app}). For GRB~090102, the presence of a large
  scale field is strongly supported by the detection of $\sim 10$\%
  optical polarization \citep{Steele:2009p5925} during a $60\,$s
  exposure shortly after the burst. It, therefore, appears that there
  is a range of magnetizations of $\sigma_0 > 0.01-0.1$ for the GRB
  ejecta at the radius of deceleration. This values are compatible
  with super-fast magnetosonic MHD models for GRBs.

\item The reverse shock slows down and compresses the shell.  As a
  result, during the RS crossing the magnetic energy of the ejecta is
  actually increased \citep[][MGA09]{Zhang:2005ts}.  Subsequently, the
  bulk of the magnetic energy is diluted because of a rarefaction that
  crosses the ejecta (produced when the RS reaches the innermost
  radial edge of the ejected shell). The remaining ($\sim$10\%) of the
  magnetic energy decays slowly with time (see Fig.~1). If a
  substantial fraction of this energy is radiated away because of
  dissipative MHD processes \citep{Giannios:2006hb}, it may account
  for the X-ray flares observed minutes to hours after many GRBs
  \citep{Burrows:2005kk}.

\item The bulk of the magnetic energy is transferred into the shocked
  ISM shortly after the end of the burst. Only at this point, the
  blast wave establishes the well known BM self-similar behavior. A
  flat or rising light curve until well after the GRB is over at
  high-energies is the signature imprinted on the FS emission by the
  aforementioned energy transfer. An initially flat high-energy light
  curve may be the signature of $\sigma_0\sim 1$ ejecta while a rising
  one comes from smaller magnetization (assuming always a constant
  density ISM).  When the energy transfer is completed, an achromatic
  break takes place. After the break, the light curve declines as
  expected from the BM solution.

\item It has been proposed \citep{Kumar:11p7654,Ghisellini:2010p7532}
  that the GeV emission observed by the LAT detector on FERMI in
  several bursts is result of the FS. The model can account for GeV
  observations shortly after the end of the burst and and for the late
  optical and X-ray data. In its simplest form, however, the FS model
  cannot explain the peak of the GeV emission at times
  $t_{peak}<t_{\rm GRB}$ seen in GRB~080916C, 090510 and likely in
  090902B (since, as we have shown here, the energy transfer to the FS
  results in $t_{peak}\simmore t_{\rm GRB}$). On the other hand, from
  the \citet{Ghisellini:2010p7532} sample of several LAT bursts, at
  least three GRBs (090323, 090626, and 091003) show a rather flat GeV
  light curve that steepens only after the MeV emission (i.e. the
  traditional GRB) is over. Though more detailed modeling is needed,
  the emission from these bursts appears compatible with coming from
  the early FS driven by ejecta with $\sigma_0\ge 1$ in agreement with
  inferences we have made from the lack of observed of RS emission.

\end{enumerate}

The results outlined so far suggest that to explain the paucity of
observed optical flashes one needs to argue that most of the observed
bursts posses substantial magnetizations (point (ii) above). However,
our models are sensitive to a number of parameters, among which,
$\epsilon_B$ is probably the most crucial one. The value that we use
as representative in our models ($\epsilon_B \sim 5\times 10^{-3}$;
Sect.~\ref{sec:lightcurves}) results in powerful optical flashes
even for the $\sigma_0=0$ model.  Values of $\epsilon_B\ll
5\times10^{-3}$ reduce the strength of the RS emission. We find that
for $\epsilon_B\simless 10^{-6}-10^{-5}$ the RS emission falls below
that of the FS, for all our purely hydrodynamic models
($\sigma_0=0$). Therefore, even in fireball models one may argue that
the optical emission of the RS is suppressed if mildly relativistic
shocks are not able to efficiently generate stochastic magnetic
fields.

Another issue that has to be taken into consideration is the fact that
we are modeling the ejecta as initially being cold. Moderately hot or
{\it warm} ejecta (i.e., in which $P/\rho c^2\simless 1$) may result,
e.g., if the conversion of internal to kinetic energy is slower than
predicted by the standard fireball model (which might not be unlikely
by looking at the numerical models of, e.g.,
\citealt{Aloy:2000ad,Aloy:2005zp}), or if internal shocks heat up the 
ejecta shortly before the reverse shock crossing. In such scenario, a weaker
RS may be envisioned, since it is more difficult to shock warm than cold
matter. Thus, we may not exclude the possibility that the absence of a
clear RS signature in many GRBs arises (assuming that the ejecta is
not magnetized) from the fact that the ejecta is warm by the time the
afterglow emission takes place.

\section*{Acknowledgments}
MAA is a Ram\'on y Cajal Fellow of the Spanish Ministry of Education
and Science. DG is a Lyman Spitzer, Jr Fellow of the Department of
Astrophysical sciences of Princeton University. We acknowledge the
support from the Spanish Ministry of Education and Science through
grants AYA2007-67626-C03-01 and CSD2007-00050 and the grant
PROMETEO/2009/103 of the Valencian Conselleria d'Educaci\'o. The
authors thankfully acknowledge the computer resources, technical
expertise and assistance provided by the Barcelona Supercomputing
Center - Centro Nacional de Supercomputaci\'on.

\bibliographystyle{mn2e}
\bibliography{mulaft}

\end{document}